# Fabrication of plasmonic surface relief gratings for the application of band-pass filter in UV-Visible spectral range

Sudheer[1,2,*], S. Porwal[3], S. Bhartiya[4], C. Mukharjee[5], P. Tiwari[1,2], T. K. Sharma[1,3], V. N. Rai[1,2], and A. K. Srivastava[1,2]

[1]*Homi Bhabha National Institute, Mumbai, Maharashtra 400094, India.*

[2] *Synchrotrons Utilization Section, Raja Ramanna Centre for Advanced Technology Indore, Madhya Pradesh 452013, India.*

[3]*Semiconductor Materials Laboratory, Raja Ramanna Centre for Advanced Technology Indore, Madhya Pradesh 452013, India.*

[4]*Laser Materials Section, Raja Ramanna Centre for Advanced Technology Indore, Madhya Pradesh 452013, India.*

[5]*Optical Coatings Laboratory, Advanced Laser & Optics Division, Raja Ramanna Centre for Advanced Technology, Indore, Madhya Pradesh 452013, India.*

*\*Corresponding author. E-mail: sudheer@rrcat.gov.in; sudheer.rrcat@gmail.com*

**Abstract:** The measured experimental results of optical diffraction of 10, 5 and 3.4 μm period plasmonic surface relief grating are presented for the application of band-pass filter in visible spectral range. Conventional scanning electron microscopic (SEM) is used to fabricate the grating structures on the silver halide based film (substrate) by exposing the electron beam in raster scan fashion. Morphological characterization of the gratings is performed by atomic force microscopy (AFM) shows that the period, height and profile depends on the line per frame, beam spot, single line dwell time, beam current, and accelerating voltage of the electron beam. Optical transmission spectra of 10 μm period grating shows a well-defined localized surface plasmon resonance (LSPR) dip at ~366 nm wavelength corresponding to gelatin embedded silver nanoparticles of the grating structure. As the period of the grating reduces LSPR dip becomes prominent. The maximum first order diffraction efficiency (DE) and bandwidth for 10 μm period grating are observed as 4% and 400 nm in 350 nm to 800 nm wavelength range respectively. The DE and bandwidth are reduced up to 0.03% and 100 nm for 3.4 μm period grating. The profile of DE is significantly flat within the diffraction bandwidth for each of the gratings. An assessment of the particular role of LSPR absorption and varied grating period in the development of the profile of first order DE v/s wavelength are studied. Fabrication of such nano-scale structures in a large area using conventional SEM and silver halide based films may provide the simple and efficient technique for various optical devices applications.

The diffractive gratings are of the particular interest as a spectroscopic analytical tool among the various diffractive optical elements to study the fundamental and engineering science.[1-2] Among the two types of gratings (reflection and transmission), generally, a reflection grating is commonly used for most of the optical applications.[3] Recently, transmission gratings are explored for achieving 100% diffraction efficiency (DE) along with tunable spectral bandwidth, which are less sensitive to the optical misalignment.[4] In transmission type, volume phase hologram (VPH) gratings are based on the periodic modulation in refractive index (maximum possible value <0.15) of grating materials. For higher DE large modulation in the refractive index of the grating medium has to be produced, which is a relatively difficult task. In this case large grating thickness (~5-10 μm) is also required that degrades the wavefront quality because of thickness variation (~5-10%) depends on the film coating techniques. Moreover, 100% DE of VPH gratings is possible only in a narrow wavelength bandwidth.[5] That is why focus moved on other transmission type i.e. surface relief transmission grating (SRTG). In GRTG periodical modulation in surface morphology of a thin layer (<1μm) effectively produces large periodic contrast in the refractive index (>0.25) of the grating material. The low variation in grating thickness and high periodic variation in refractive index of SRTG full fills the basic criteria of optical diffraction along with the good performance. For fabrication of SRTG structures, exposure-based techniques attract much attention compared to ruling based techniques because of the absence of ghost light and the presence of low stray light. In the optical beam lithography (OBL), the contrast of exposure produced by the interference patterned of two parallel laser beams become low because of overlapping of scattered light generated from dust, digs, and scratches of the used optical components that results in quality reduction of fabricated structures. Out of various exposure-based techniques, electron beam lithography (EBL) is renowned for the fabrication of good quality small features but at the same time it's inconspicuous because of time consuming, costly and low throughput (small area of fabrication). However, the requirement of simple and fast fabrication technique along with large fabrication area, turns the attention towards the EBL by conventional scanning electron microscope (SEM).[6] Moreover, incorporation of metal particles in the structures opens up the new flexibility in the fabrication along with the applicability in various technologies including plasmonic waveguide, plasmonic filter, magnetic memory arrays, plasmonic lenses in terms of manipulation and confinement of light and electrons at nano-scale.[7-9] J. Cesario et. al employed the electromagnetic coupling between metal nanoparticle grating and metallic surface to modulate the extinction spectra in visible range.[10] Particularity, various authors have used the nanoparticles to fabricate the plasmonic absorption gratings for the application of tuned diffraction response,[11] optical switches[12-16] and surface enhanced Raman spectroscopy (SERS).[6,17] These absorption gratings are fabricated by producing a periodic modulation in the optical absorbance. The variation in the size as well as number density of silver nano particles in the grating strips effectively lead to special periodic variation in the optical absorbance that result to form an absorption grating.

In the metal nanoparticle conduction band electrons are able to oscillate collectively once the dimension of metal nanoparticle is smaller than the incident wavelength of light that results in the extraordinarily transmission, absorption, and scattering.[18] This unique optical behavior of nanoparticles in contrast to bulk material is attributed to localized surface plasmon resonance (LSPR).

Silver halide is one of the outstanding photographic material for recording the images (holograms) via light and electron beam exposure whereas polyethylene terephthalate (PET) is regarded as light weight and low cost visible transparent substrate, which is suitable as a supporting base material for transmission kind of structures. Recently, fabrication of plasmonic surface relief transmission gratings (PSRTGs) of silver nanoparticle have been performed using conventional SEM and PET supported silver halide film for the application of SERS substrate, where the enhancement in the Raman signal is observed as the period of the grating reduced.[6] Variation in the diffraction efficiency (DE) of PSRTGs fabricated at different accelerating voltages have also been reported earlier.[19]

The main aim of the present study is to fabricate PSRTGs of variable periodicity using conventional SEM and silver halide based transmission electron microscope (TEM) film as well as to study the effect of grating periodicity on the bandwidth and the diffraction efficiency (DE) of the gratings. The LSPR absorbance of silver nanoparticles present in the gelatin matrix of grating material is correlated with its morphology, bandwidth and DE.

In the experimental procedure, the silver halide based commercially available electron microscope film (Kodak 4489) was used as a bare substrate for the fabrication of PSRTGs.[20] These films are specially engineered to produce the high-resolution images in transmission electron microscopes (TEM). The bare substrate consists four layers. The base layer is made of PET of ~278 μm thickness, which is over coated with the layer of gelatin embedded silver halide nanoparticles (called emulsion) of ~3.5 μm thickness. This emulsion layer is protected with a top pure gelatin layer of ~100 nm thickness. The backside of the PET is coated with a carbon layer (bottom layer) of ~100 nm thickness, which reduces the back scattering of the electron during EBL process. Fabrication of PSRTGs is performed using a conventional SEM by electron beam exposure on the substrate in raster scan fashion with 30 keV beam energy. The other operational parameters of exposure beam like spot size, beam current, single line dwell time, and magnification are optimized as 1 (corresponds to ~10 nm), 65 mA, 40 ms, and 20× respectively. The area of fabrication is controlled by variation in the magnification. The variation in the line per frame during exposure at fixed magnification offers flexibility to change the grating period. In order to change the periodicity of grating in the present fabrication 484, 968, and 1452 lines per frame are used to exposed the substrate.[6] The technique involved in the measurement of first order DE have been described in detail elsewhere and outlined below.[19] The experimental set-up consist of a monochromatic collimated beam of light ~2 mm in diameter. This beam falls on the grating structures perpendicularly from the air-PET interface side, which produces a diffraction pattern towards the grating-air interface. Thereafter, only the first order light is selected by an aperture and its intensity is recorded from 350 nm to 800 nm wavelengths using a Si detector having ~5 mm diameter entrance. In order to enhance the signal to noise ratio, a phase sensitive detection scheme is implemented with the help of a mechanical chopper and SRS 830 Lock-in amplifier interfaced with the computer for automatic data recording. The DE of the first order is defined as the ratio of 1st order intensity ($I_{+1}$) to the incoming intensity ($I_0$) at each wavelength. The $I_0$ is recorded by replacing the grating structure by a blank PET in the same spectral range which includes all the losses of optical components and bare PET substrate.

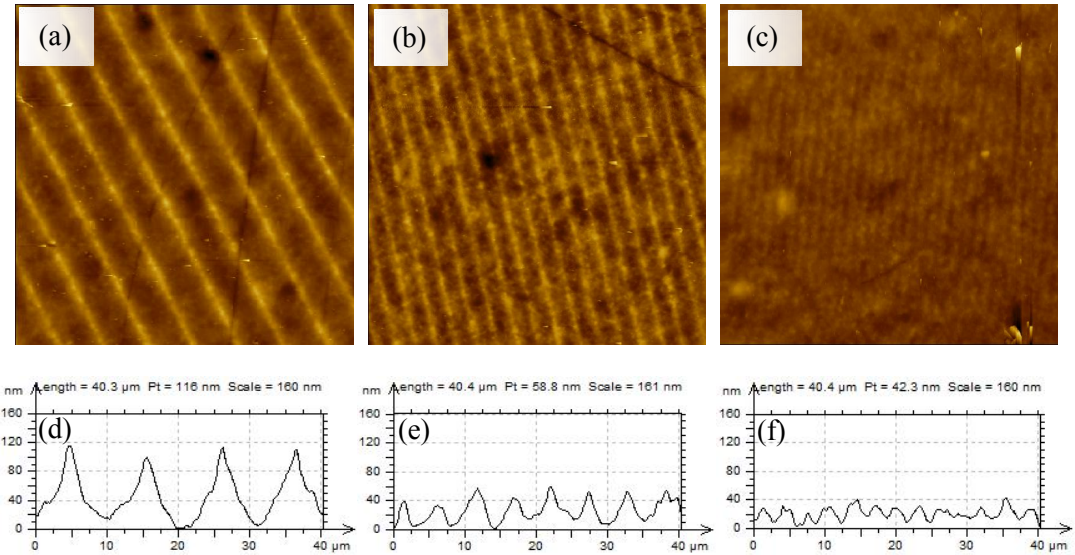

FIG.1 *AFM images of plasmonic surface relief grating structures of different period (fabricated at 30 kV electron beam accelerating voltage) along with line scan profiles of the gratings (a, d) 10 µm, (b,e) 5 µm and (c,f) 3.4 µm. Line scan is taken perpendicular to the grating axis. Scanning area for each 2-dimentional AFM images is 90µm by 90µm.*

In the results section, first of all, the bare substrate has been characterized. The thickness of the silver halide emulsion layer is measured as ~3.5 µm by cross-section field emission electron microscope (FESEM) imaging. The platelet-shaped mono-dispersed silver halide nanoparticles are found uniformly suspended in the gelatin matrix of the emulsion layer. The dimensions of the silver halide grains were measured as ~450 nm in length and ~120 nm in width. The presence of Ag and Br in bare substrate are also confirmed by energy dispersive spectroscopy (EDS).[6,19]

The morphology of the grating structures is imaged by atomic force microscope (AFM). The raster scan exposure of electron beam is clearly recorded as equally spaced sharper parallel lines on the silver halide based film as shown in 2-dimentional AFM images (Figs. 1(a)-1(c)). To investigate the period, depth and profile of the fabricated PSRTGs, line scan is obtained from the 2-dimentional AFM images. Figs. 1(d)-1(f) show line scan of the fabricated grating structures corresponding to the 484, 968 and 1452 lines per frame. The period of the fabricated gratings are found as 10, 5 and 3.4 µm corresponding to 484, 968 and 1452 lines per frame. It is obvious that at a fixed magnification (20×) exposure area on the substrate remains constant (12.5 mm$^2$) for any number of exposure lines per frame in a single scan. In that case, an increase in the lines per frame of the exposure beam decreases the period of the fabricated grating. The depth of the fabricated gratings was also measured from the line scan profile as ~110 nm, ~50 nm and ~30 nm for 484, 968 and 1452 lines per frame respectively. The reduction in depth of the grating for lower period occurred due to overlapping of the electron beam that produces a bottom layer of silver nanoparticles below the grating, which effectively reduces the absolute depth of the grating. However, the number density in the grating (silver present in the grating as well as in bottom layer) is increased for lower period grating structures. It has been shown by energy dispersive spectroscopy (EDS) techniques that all the exposed silver halide grains in the emulsion converts into silver grains and unexposed grains dissolve in fixer solution at the end of fabrication process. The particle size of the silver nanoparticles present in the fabricated grating has also been calculated using the X-ray diffraction (XRD) technique as ~15-20 nm.[6]

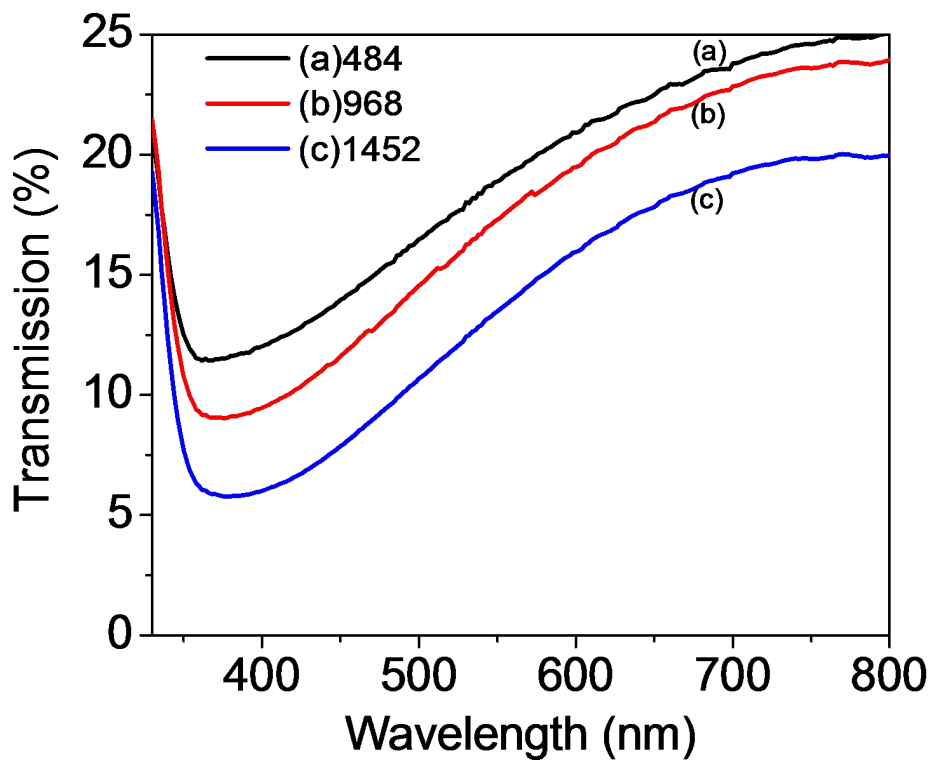

*FIG.2  Absorption spectra of the plasmonic surface relief grating structures of different period fabricated at 30 kV electron beam accelerating voltage.*

Fig. 2 shows the transmission spectra of fabricated PSRTG structures. The dip ~366 nm in transmission spectra for 10 μm period grating is ascribed due to the LSPR absorption of the silver nanoparticles embedded in the grating structure.[6,19] The dip in the transmission spectra becomes prominent as the grating period reduces. This is because of an increase in the number density of silver nanoparticles as discussed above from the AFM line scan. It is noted the transmission dip position slightly shift toward higher wavelength side (red shift) and broadening in the LSPR profile increases with decrease in the grating period. Moreover LSPR strength in terms reduction in transmission at dip position also increases. Such change are expected due to increase in the particle-particle interaction as a result of increases in the number density of silver nano particles with decrease in the period of gratings.

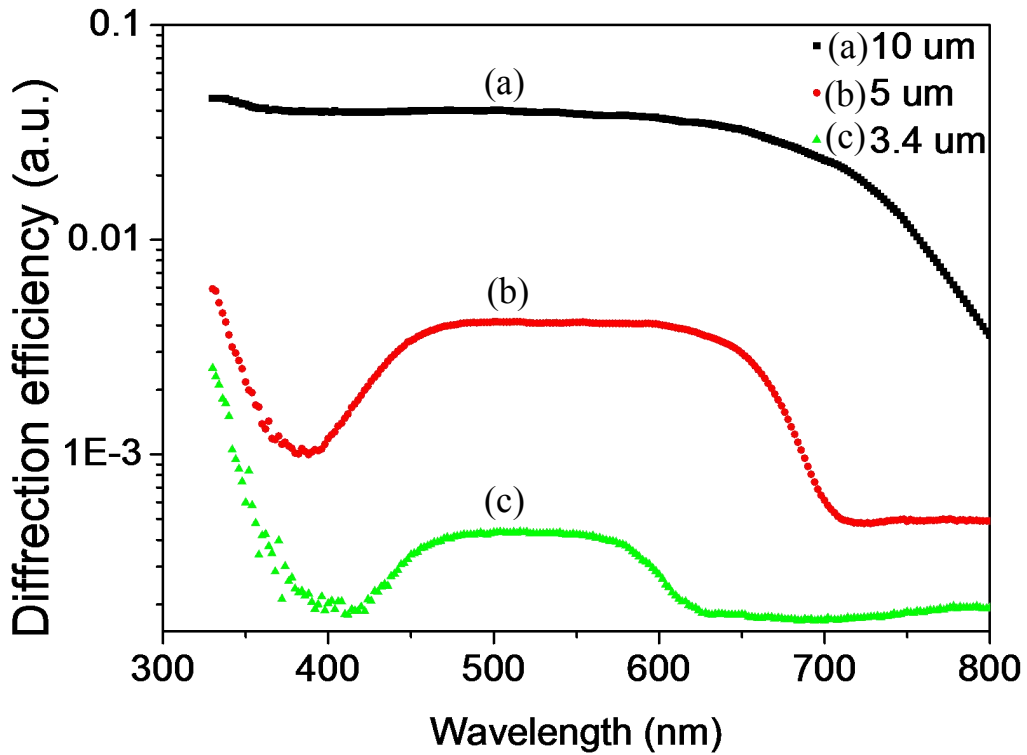

*FIG. 3 Variation in the experimentally measured first order diffraction efficiency and bandwidth for plasmonic surface relief grating structures of different period fabricated at 30 kV electron beam accelerating voltage.*

The investigation of optical diffraction properties of all the fabricated PSRTGs has been performed. Fig. 3 shows the experimentally measured first order DE with respect to the wavelengths for the gratings of 10, 5 and 3.4 μm period from 350 to 800 nm visible spectral ranges. The maximum first order DE of ~4% is observed for 10 μm period grating in 350 to 800 nm spectral range (bandwidth~450 nm). Moreover, the nature of the DE is remarkably flat over the 350 to 650 nm (~flatten bandwidth 300 nm) spectral range and dropped significantly at the higher wavelength side. As the grating period decreases from 10 μm to 5 μm, the magnitude of DE reduces almost by one order. In-addition, suppression in the bandwidth also occurs from both the sides of the spectral wavelength that makes DE profile narrower as ~300 nm (~400 to ~700 nm). Nevertheless, flatten nature of DE profile remains same in the suppressed bandwidth. The flat spectral bandwidth reduced to ~200 nm (from ~450 to ~650 nm). In this case 3.4 μm period grating the magnitude of DE again reduces by nearly one order of magnitude than the 5 μm period grating. Similarly band width also shows corresponding decrease from 300 to 200 nm, but the profile remained nearly flat. The comparison of Fig. 2 and Fig. 3 shows that the diffraction efficiency as well as the band width of the gratings decreases with an increase in the number of lines in the grating (decrease in the period). This may be either due to decrease in the height of grating or due to increase in the number density of silver nanoparticles in the grating as the period of grating decreases.

The occurrence of flat nature in the DE profiles, reduction in DE magnitude and narrowing in the bandwidth can be understood qualitatively in terms two effects: (1) decrease in the grating period and its height (2) LSPR based increase in the absorption of silver nanoparticle embedded in the grating material (gelatin matrix). In fact fabricated gratings are the "absorption gratings" in which periodic variation in the height of PSRTG produces a periodic variation in the extent of silver nanoparticles that ultimately leads the modulation of absorption coefficient ($\alpha$). It can be seen that as the grating period decreases, it accompanies a relative decrease in the height of the grating along with an increases in the thickness of the bottom layer that leads to a reduction in the modulation amplitude of absorption coefficient ($\Delta\alpha_1$) and an enhancement in average absorption coefficient ($\alpha_{ave}$). The decrease in the $\Delta\alpha_1$ seems to be playing important role in reducing the bandwidth of diffraction profile of the grating, whereas increase in the $\alpha_{ave}$ produces the significant absorption that ultimately leads to the reduction in the DE of the grating. This increased absorption may be associated with the increased scattering DE of the grating. It is reported earlier about the SRTG (dielectric) that spectral band width is directly proportional to the period and inversely proportional to the thickness of the grating. In the present case thickness effect is found opposite or may be negligible. Results indicate that decreasing the thickness of uniform layer of silver nanoparticle below grating may increases the efficiency of the grating manifold. However, full theoretical reproduction of DE profiles required rigorous calculation of Raman-Nath diffraction theory. It seems that DE enhancement and bandwidth shortening/broadening can be tuned by variation in the thickness bottom absorbing layer, grating height and grating period by optimizing the EBL technique. Finally, this is a very easy, low cost and less time consuming technique providing a large band width with a fate DE profile.

In summary, tunable optical diffraction efficiency (DE) along with significantly flat diffraction band for plasmonic surface relief grating (PSRG) have been observed due to variation in the grating period and amount of silver nanoparticles. These PSRGs of period 10, 5, and 3.4 μm are fabricated by the conventional scanning electron microscope (SEM) on silver halide based electron microscope film by single exposure of the electron beam in raster scan fashion. Area of fabrication (~12.5 mm$^2$) and period of the PSRGs was controlled by magnification (20×) and number of scanning line per frame respectively. First order DE and bandwidth are measurement as 4% and 400 nm in the visible range for the 10 μm period of the grating. The DE and bandwidth are reduced to 0.03 % and 100 nm for 3.4 μm. The reduction in DE and narrowing in bandwidth are found correlated with the localized surface plasmon resonance of silver nanoparticle embedded in the grating itself and its period prospectively. The natures of DE profile found significantly flatten for each of the structure within the bandwidth. Such controlling the DE and bandwidth of (PSRG) open up the possibility of potential applications in various optical devices like a band-pass filter for visible spectral range.

**Acknowledgments**

One of the authors Sudheer acknowledges the Homi Bhabha National Institute (HBNI) for providing financial assistance. Authors are thankful to Mr. Mahendra Babu for providing help in the fabrication of PNG structures.